\documentclass[letter]{aa} 

\usepackage{graphicx}
\usepackage{txfonts}
\usepackage{hyperref}  
\hypersetup{colorlinks=true,linkcolor=[rgb]{1.,0.2,0.2},citecolor=[rgb]{0.1,0.4,1.},filecolor=[rgb]{0.7,0.2,0.2},urlcolor=[rgb]{0.7,0.2,0.2}}

\definecolor{blue}{rgb}{0., 0., 1}

\newcommand{\oiii}{[\textrm{O}\textsc{iii}]}

\newcommand{\oiiiv}{[\textrm{O}\textsc{iii}]\ensuremath{\lambda5007}}

\newcommand{\oiiialone}{[\textrm{O}\textsc{iii}]}

\newcommand{\oiiidoublam}{[\textrm{O}\textsc{iii}]\ensuremath{\lambda\lambda4959,5007}}
 
\newcommand{\ha}{\ifmmode {\rm H}\alpha \else H$\alpha$\fi}
\newcommand{\halam}{\ifmmode {\rm H}\alpha \lambda6563 \else H$\alpha$ $\lambda$6563 \fi}
\newcommand{\hb}{\ifmmode {\rm H}\beta \else H$\beta$\fi}
\newcommand{\hg}{\ifmmode {\rm H}\gamma \else H$\gamma$\fi}
\newcommand{\hblam}{\ifmmode {\rm H}\beta \lambda4861 \else H$\beta$ $\lambda$4861 \fi}
\newcommand{\lya}{\ifmmode {\rm Ly}\alpha \else Ly$\alpha$\fi}
\newcommand{\pg}{\ifmmode {\rm P}\gamma \else Pa$\gamma$\fi}
\newcommand{\lyb}{\ifmmode {\rm Ly}\beta \else Ly$\beta$\fi}
\newcommand{\lyg}{\ifmmode {\rm Ly}\gamma \else Ly$\gamma$\fi}

\newcommand{\heii}{\textrm{He}\textsc{ii}\ensuremath{\lambda1640}}

\newcommand{\flyc}{\ifmmode  \mathrm{f}_\mathrm{esc}\mathrm{(LyC)} \else $\mathrm{f}_\mathrm{esc}\mathrm{(LyC)}$\fi}

\def\kms{km s$^{-1}$}

\def\ergs{\ifmmode \mathrm{erg\hspace{1mm}s}^{-1} \else erg s$^{-1}$\fi}
\def\ergscm{erg s$^{-1}$ cm$^{-2}$}
\def\micron{\ifmmode \mu\mathrm{m} \else $\mu$m\fi}
\def\msun{\ifmmode \mathrm{M}_{\odot} \else M$_{\odot}$\fi}
\def\msunyr{\ifmmode \mathrm{M}_{\odot} \hspace{1mm}{\rm yr}^{-1} \else $\mathrm{M}_{\odot}$ yr$^{-1}$\fi}
\def\zsun{\ifmmode Z_{\odot} \else Z$_{\odot}$\fi}
\def\lsun{\ifmmode L_{\odot} \else L$_{\odot}$\fi}
\def\mstar{\ifmmode \mathrm{M}_{\star} \else M$_{\star}$\fi}
\newcommand{\JWST}{\textrm{JWST}}
\newcommand{\HST}{\textrm{HST}}

\usepackage{orcidlink}

\begin{document}

\titlerunning{\JWST\ probes super-faint low-metallicity stellar ionizers at $z\simeq4.19$}

\title{A pristine, star-forming complex at $z=4.19$\thanks{Based on observations collected with the James Webb Space Telescope (\JWST) and Hubble Space Telescope (\HST).
These observations are associated with \JWST\ GO n.1908 (PI E. Vanzella), GTO n.1208 (CANUCS, PI Willott), and GTO n.1176 (PEARLS, PI Windhorst).}}

\authorrunning{Eros Vanzella et al.}
\author{
E.~Vanzella\inst{\ref{inafbo}}\fnmsep\thanks{E-mail: \href{mailto:eros.vanzella@inaf.it}{eros.vanzella@inaf.it}}$^{\orcidlink{0000-0002-5057-135X}}$ \and
M.~Messa\inst{\ref{inafbo}}$^{\orcidlink{0000-0003-1427-2456}}$ \and
A.~Zanella\inst{\ref{inafbo}}$^{\orcidlink{0000-0001-8600-7008}}$\and
A.~Bolamperti\inst{\ref{mpa}}$^{\orcidlink{0000-0001-5976-9728}}$\and
M.~Castellano\inst{\ref{inafroma}}$^{\orcidlink{0000-0001-9875-8263}}$ \and
F.~Loiacono\inst{\ref{inafbo}}$^{\orcidlink{0000-0002-8858-6784}}$ \and
P.~Bergamini\inst{\ref{inafbo}}$^{\orcidlink{0000-0003-1383-9414}}$ \and
G.~Roberts~Borsani\inst{\ref{UnivCollegeLondon}}$^{\orcidlink{0000-0002-4140-1367}}$ \and 
A.~Adamo\inst{\ref{univstock}}$^{\orcidlink{0000-0002-8192-8091}}$\and
A.~Fontana\inst{\ref{inafroma}}$^{\orcidlink{0000-0001-9875-8263}}$ \and
T.~Treu \inst{\ref{UCLA}}$^{\orcidlink{0000-0002-8460-0390}}$
F.~Calura\inst{\ref{inafbo}}$^{\orcidlink{0000-0002-6175-0871}}$\and
C.~Grillo \inst{\ref{unimi},\ref{inafiasf}}$^{\orcidlink{0000-0002-5926-7143}}$\and
M.~Lombardi\inst{\ref{unimi}}$^{\orcidlink{0000-0002-3336-4965}}$ \and
P.~Rosati \inst{\ref{unife},\ref{inafbo}}$^{\orcidlink{0000-0002-6813-0632}}$\and
R.~Gilli\inst{\ref{inafbo}}$^{\orcidlink{0000-0001-8121-6177}}$\and
M.~Meneghetti \inst{\ref{inafbo}}$^{\orcidlink{0000-0003-1225-7084}}$
}
\institute{
INAF -- OAS, Osservatorio di Astrofisica e Scienza dello Spazio di Bologna, via Gobetti 93/3, I-40129 Bologna, Italy \label{inafbo} 
\and
Max-Planck-Institut f\"ur Astrophysik, Karl-Schwarzschild-Str. 1, D-85748 Garching, Germany\label{mpa}
\and
INAF -- Osservatorio Astronomico di Roma, Via Frascati 33, 00078 Monteporzio Catone, Rome, Italy\label{inafroma}
\and
Department of Physics \& Astronomy, University College London, London, WC1E 6BT, UK\label{UnivCollegeLondon}
\and
Department of Astronomy, Oskar Klein Centre, Stockholm University, AlbaNova University Centre, SE-106 91, Sweden\label{univstock}
\and
Department of Physics and Astronomy, University of California, Los Angeles, 430 Portola Plaza, Los Angeles, CA 90095, USA\label{UCLA}
\and
Dipartimento di Fisica, Università degli Studi di Milano, Via Celoria 16, I-20133 Milano, Italy\label{unimi}
\and
INAF -- IASF Milano, via A. Corti 12, I-20133 Milano, Italy\label{inafiasf}
\and
Dipartimento di Fisica e Scienze della Terra, Università degli Studi di Ferrara, Via Saragat 1, I-44122 Ferrara, Italy\label{unife}
}

\date{} 

 
\abstract 
{
We report the discovery of a faint (M$_{\rm 1700} \simeq -12.2$), oxygen-deficient, strongly lensed ionizing source \textemdash\ dubbed Lensed And Pristine 2 (LAP2) \textemdash\ at a spectroscopic redshift of $z=4.19$.
LAP2 appears to be isolated and lies very close to the caustic produced by the lensing galaxy cluster Abell 2744. It was observed with the \textit{\emph{James Webb Space Telescope}} (\JWST) NIRSpec MSA in prism mode as part of the UNCOVER program. The NIRSpec spectrum reveals prominent \lya\ ($7.1\sigma$), clear \ha\ ($6.2\sigma$), tentative \hb\  ($\simeq 2.8\sigma$) emissions and no detectable \oiiidoublam~($\sim 7$ times fainter than \ha). The inferred \oiiialone\ $2\sigma$ upper limit corresponds to an R3 index $<0.85$ (assuming the \ha/\hb~=~2.86 case~B recombination ratio), which, under high-ionization conditions, implies a metallicity of $Z < 6 \times 10^{-3}~Z_{\odot}$.
The combination of faint ultraviolet luminosity, a large rest-frame \ha\ equivalent width ($\simeq 650$~\AA), and an extremely compact size ($<10$~pc) suggests that LAP2 is being caught in an early, pristine formation phase consistent with an instantaneous-burst scenario, with an estimated stellar mass of at most a few $\times 10^4$~M$_\odot$. Deep VLT/MUSE observations further reveal copious \lya\ emission forming an arclet that straddles the critical line. 
LAP2 joins the rare class of extremely metal-poor star-forming complexes that the \JWST\ has started to unveil at redshifts $3 - 7$, and it provides a glimpse into a still very poorly explored low-luminosity regime.
}
\keywords{galaxies: high-redshift -- galaxies: star formation -- stars: Population III -- gravitational lensing: strong.}
   \maketitle

\section{Introduction}
\label{sect:intro}

The direct observation of PopulationIII (Pop~III) star formation remains a challenging task \citep[e.g.,][]{Katz2023} and typically requires the aid of gravitational lensing, which boosts the intrinsically faint continuum and line emission \citep[e.g.,][]{zack2015}. Lensing also enhances spatial contrast, making it possible to distinguish individual — and potentially isolated — stellar pockets. Depending on the magnification, this enables investigations on scales of tens of parsecs, and in some cases down to parsec-sized regions \citep[e.g.,][]{vanz_popiii,adamo2024,Messa_D1T1_2024}.
The advent of the \JWST\ facility, with its greatly improved angular resolution and extended wavelength coverage into the near-infrared, has opened the way for targeted searches for Pop~III galaxy candidates \citep[e.g.,][]{Trussler2023,Fujimoto2025} and, potentially, for their direct detection. The recent discovery of two oxygen-deficient, high-redshift sources — AMORE-B at $z=5.725$ \citep{Morishita2025} and CR3 at $z=3.19$ \citep{Cai2025} — adds to the previously identified lensed and pristine object at $z=6.63$ (LAP1), initially serendipitously discovered  with VLT/MUSE \citep{vanz_popiii} and subsequently followed up with the \JWST\ \citep{Vanzella2023_lap1,Nakajima2025}. While the detailed physical properties of these systems are still under investigation, the apparent deficit of oxygen lines in the rest-frame optical suggests gas-phase metallicities below 1\% of the solar value, and re-opens the issue of prolonged pristine star formation down to relatively low redshift: $z\sim 3-5$ \citep[e.g.,][]{Liu_Bromm_endPopIII2020, Hegde2025}.
Finding very low-metallicity, low-mass sources at these redshifts is important because they provide a rare glimpse of galaxies under conditions similar to those of the early Universe, helping us understand how the first generations of stars (Pop~III/II) enriched the cosmos with heavier elements (\citealt[][]{venditti2023, maio2016}; see also \citealt{Rusta2025}). At the same time, they allow us to probe metallicity at very low stellar masses, which is particularly relevant for constraining the mass–metallicity relation down to unexplored mass limits \citep[][]{Maiolino_Mannucci2019}.
The rarity of these objects in deep-field surveys underscores the importance of strong gravitational lensing to bring them within reach of current instrumentation \citep[e.g.,][]{zack2015}.

In this work, we present \JWST/NIRSpec MSA observations of an extremely faint ionizing source, first identified with VLT/MUSE as an arclet straddling the critical line at $z=4.19$ (1.5 Gyr after the Big Bang) and included among the spectroscopically confirmed multiple images in the \citet{Bergamini2023} lens model. The system lies behind the Hubble Frontier Fields cluster Abell~2744 \citep{Lotz_2017HFF}, which strongly lenses the source and reveals its oxygen-deficient nature at $z=4.19$. We assumed a flat cosmology with $\Omega_{\rm M}=0.3$, $\Omega_{\rm \Lambda}=0.7$, and $H_{0}=70,{\rm km},{\rm s}^{-1},{\rm Mpc}^{-1}$. Magnitudes are in the AB system \citep{Oke_1983}, $m_{\rm AB}=23.9-2.5\log(f_\nu/\mu{\rm Jy})$.

\begin{figure}
\center
 \includegraphics[width=0.95\columnwidth]{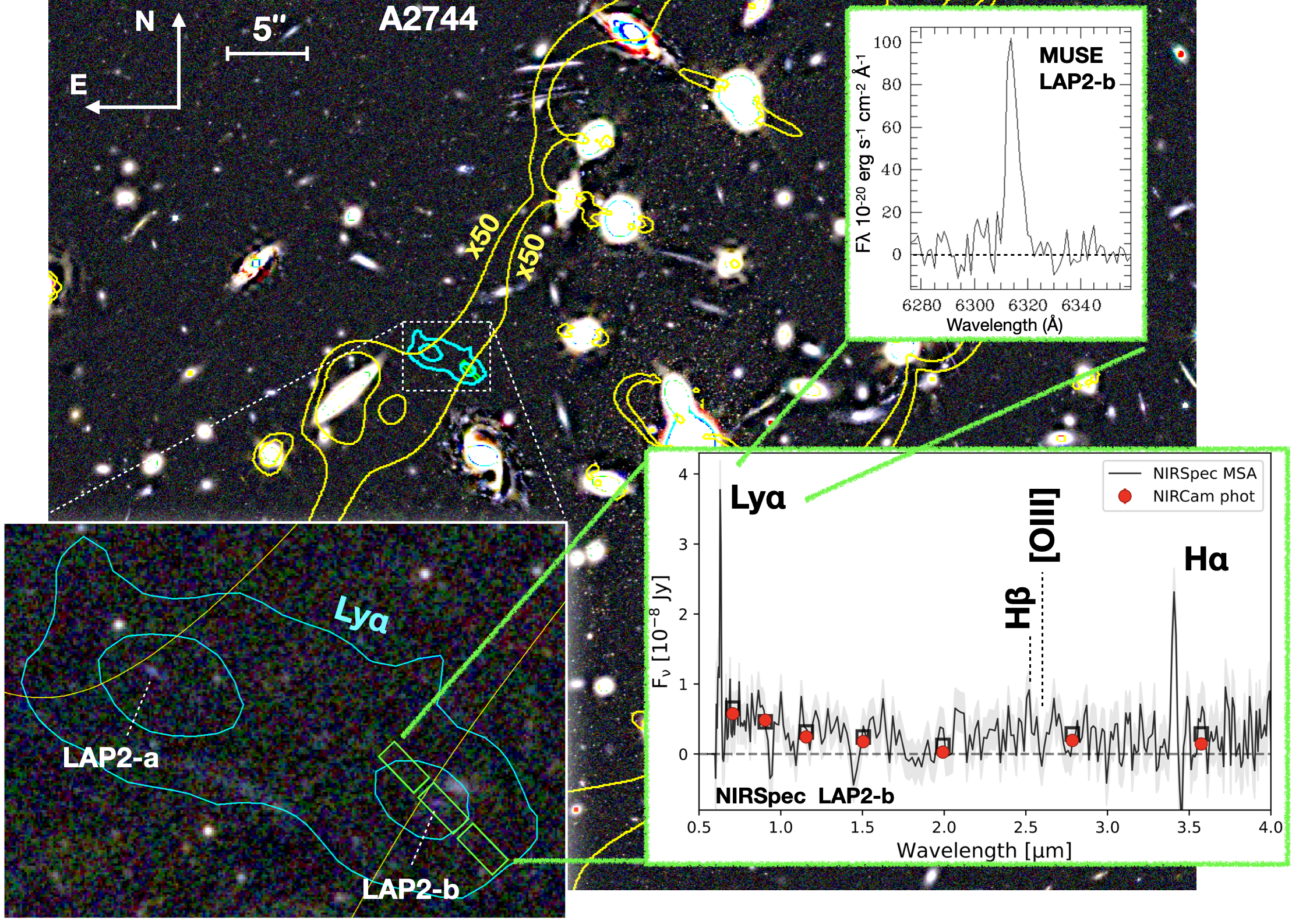}
 \caption{Schematic view of \lya\ emitter at $z=4.189$ amplified by the galaxy cluster A2744 (background color image, F090W, F150W, F200W). The yellow contours mark $\mu=50$ at $z=4.19$ \citep[][]{Bergamini2023}. The NIRSpec MSA pointing (green box) is shown in the bottom left inset, which zooms in on the \lya\ arclet and its multiple images, dubbed LAP2-a,b (the cyan lines outline the 3 and 8 $\sigma$ \lya\ contour from VLT/MUSE).  The upper right inset shows the VLT/MUSE \lya\ line emission of LAP2-b. The bottom right inset shows the NIRSpec prism spectrum and error of LAP2-b in F$_\nu$ units with the fluxes inferred from the spectrum convolved with NIRCam filters (open squares) and the corresponding measured NIRCam magnitudes indicated.} 
 \label{fig:main}
\end{figure}
\begin{figure*}
\center
 \includegraphics[width=0.95\textwidth]{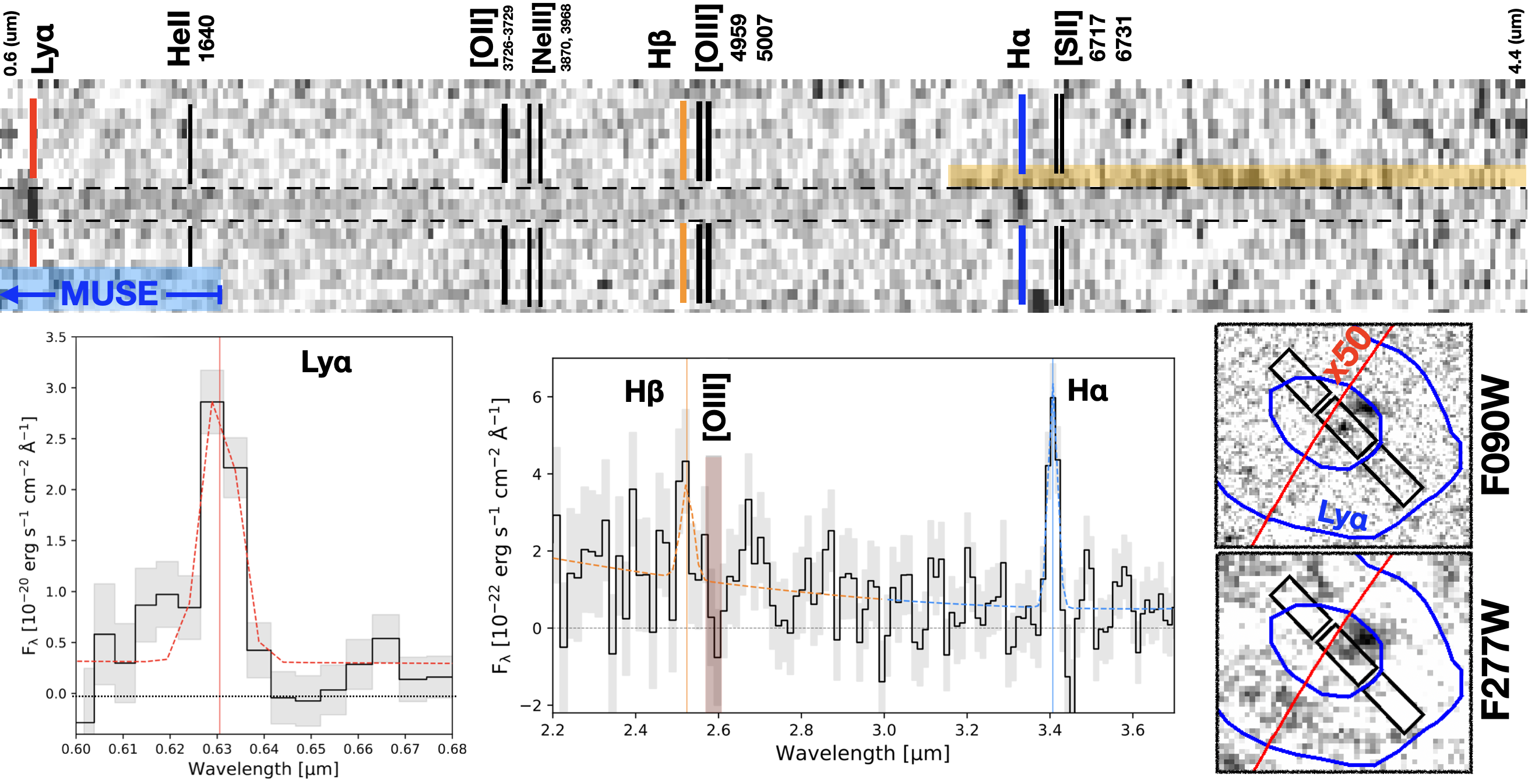}
 \caption{Overview of \JWST/NIRSpec MSA pointing on LAP2. The 2D
 spectrum is shown on the top with the more relevant lines indicated, along with the area used for the extraction of the spectrum (dashed horizontal lines). The yellow shaded area marks contamination signal arising from a nearby source, and the blue shaded area on the left marks the wavelength coverage of VLT/MUSE. The bottom left and middle insets show the line and continuum fitting (dashed colored curves). The light gray shaded area indicates the flux uncertainties, while the light red area in the bottom middle panel indicates the expected wavelength of the \oiii\ doublet. Both the 1D and 2D spectra reveal the absence of oxygen emission. The bottom right inset shows a zoomed-in view ($1.9''\times1.6 ''$) of the F090W band of LAP2-b targeted with MSA (outlined with the black boxes). The blue contours mark the VLT/MUSE \lya\ at 3 and 8 sigma, along with the locus of $\mu=50$ (red line). 
 }
 \label{fig:line}
\end{figure*}
%


\section{\JWST\ observations and analysis} \label{sec:imaging}

The galaxy cluster 
A2744 has been widely observed with \JWST/NIRCam, NIRSpec, and NIRISS since Cycle~1, including GLASS ERS \citep[][]{Treu_GLASS_2022}, UNCOVER \citep[][]{Price_UNCOVER_2025}, and MEGASCIENCE \citep[][]{Suess_2024}. We used
\JWST/NIRSpec GO~2561 data (PI: Labbe; UNCOVER; \citealt{Price_UNCOVER_2025}) targeting LAP2-b with the MSA prism/CLEAR (Figure~\ref{fig:main}). We analyzed the reduced MAST products (obsid: 300843376), which provide flux- and wavelength-calibrated 2D spectra with associated errors, for a total exposure time of 15,756~s. We extracted the spectrum at the \lya, \ha, and continuum positions using a three-pixel window (Figure~\ref{fig:line}). Line fitting was performed with {\tt specutils}/{\tt astropy}\footnote{http://www.astropy.org}
: \ha\ was modeled with a Gaussian (centroid, full width half maximum $-$~FWHM~$-$ and
flux), while for \hb\ the centroid and width were fixed to those of \ha; the continuum was fit with a second-degree polynomial. The \ha-based redshift is $z=4.189 \pm 0.03$ (Table~\ref{tab:properties}). 
NIRCam imaging from the above programs was retrieved from the Dawn JWST Archive (DJA)\footnote{https://dawn-cph.github.io/dja/}. We adopted F090W as rest-UV ($\simeq1700$~\AA) and F277W as rest-optical ($\simeq5300$~\AA). As shown in Figure~\ref{fig:main}, LAP2 is split into two magnified images, LAP2-a and LAP2-b, with $\mu \sim 58$ and $\sim 50$, respectively \citep[][]{Bergamini2023}. We focused on LAP2-b, the NIRSpec MSA target; all values in Table~\ref{tab:properties} refer to it. Figure~\ref{fig:line} shows the 1D and 2D
spectra and NIRCam imaging.
While the NIRSpec-inferred continuum and the NIRCam photometry agree within the uncertainties overall, the long wavelength NIRCam magnitudes appear slightly fainter than NIRSpec. LAP2-b is faint, lies on elevated cluster background, and is close to a potential contaminant. This likely explains the small offset between the NIRCam and NIRSpec magnitudes redward of $2 \mu$m. However, the \oiiidoublam/\ha\ ratio discussed below is computed including the continuum in the line fitting, so the \ha\ flux is negligibly affected by contamination, leaving our main conclusions unchanged. Forthcoming NIRSpec observations of LAP2-a (the counter-image) will further mitigate this issue.
None of the lines (\lya, \hb, \ha) are spectrally resolved at prism resolution ($\simeq 4000, 4000,$ and 3000~\kms). A tentative \hb\ is detected at $\sim2.8\sigma$. The \lya\ detection recalls that from MUSE, though differences in angular resolution, potential slit losses \citep[][]{Bhagwat2025}, and proximity to the NIRSpec cutoff prevent direct flux comparison. No other rest-frame UV/optical lines are detected up to $\lambda \simeq 9600$~\AA, in particular no \heii, consistent with the MUSE spectrum extending to 1800~\AA. The \heii\ line has a $2\sigma$ EW limit of 80~\AA\ (from the MUSE flux limit and NIRCam F090W continuum). This limit remains too weak to place meaningful constraints on the Population III signature \citep[][]{POPIII_Nakajima2022}.
Deeper NIRSpec/IFU prism data are allocated in Cycle~4 (prog.~7677), covering \lya\ to \ha. 

\begin{table}
\caption{Observed and derived properties of LAP2-b.} 
\label{tab:properties}      
\centering          
\begin{tabular}{l | c }   
\hline\hline  
Quantity    & LAP2-b  \\ 
\hline
RA  & 00:14:21.74\\
DEC & $-$30:23:56.1\\
Redshift~(\ha) & $4.189 \pm 0.003 $\\
$m_\mathrm{UV}$ [F090W]  & $29.69\pm0.09$  \\
$m_\mathrm{opt}$ [F277W] & $30.68\pm0.24$  \\
$M_{\rm 1700},~M_{\rm 5300}$$^{\dag}$ & $-12.2,~-11.2$ \\
\ha\ ~~[$10^{-19}$ cgs ]       &  $1.68 \pm   0.30$ \\
\oiiiv\ ~~[$10^{-19}$ cgs]    &  $< 0.50$ \\  
\hb\ ~~[$10^{-19}$ cgs ]       &  $0.72 \pm 0.26$ \\
\lya\ ~~[$10^{-19}$ cgs ]      &  $26.4^{+3.7}_{-3.8}$ \\
\lya\ ~~[$10^{-19}$ cgs ]~(MUSE)      &  $55.0 \pm  5.8$\\
EW(\ha)~~[\AA ]$^{\star}$               &  $647_{-220}^{+302}$ \\
log$\left(\frac{\xi_{ion} {\rm [Hz/erg]}}{1-{\rm fesc}}\right)$$^{\ast}$ &  $\ge 25.13_{[\rm fesc>0]}$ \\
R3~~[$=\frac{\oiiiv}{\ha/2.86}$]  & $<0.85$ \\ 
Z(Z$_{\odot}$), 12+log(O/H)$^{\ddagger}$    & $<0.006$, $<6.5$\\
$\mu_{\rm tot}$,~$\mu_{\rm tang}$ &  $50\pm5, 20\pm3$ \\
\hline \hline
\end{tabular}
\tablefoot{Unless specified, magnitudes and fluxes are not corrected for lensing. De-lensed magnitudes were obtained by adding $2.5\log_{10}(\mu_{\rm tot})$ to the observed ones, and de-lensed fluxes by dividing by $\mu_{\rm tot}$. Total and tangential magnifications ("tot", "tang") are reported in the last row with 10\% relative statistical error \citep[][]{Bergamini2023}.  
Errors and limits are at $1\sigma$ and $2\sigma$. Line fluxes are in cgs (erg~s$^{-1}$~cm$^{-2}$).  
$^{(\dag)}$ Intrinsic absolute magnitudes after lensing correction.  
$^{(\star)}$ EW(\ha) from line flux and F277W magnitude as continuum.  
$^{(\ast)}$ Lower limit, assuming $f_{\rm esc}>0$.
$^{(\ddagger)}$ From \citet{asplund2009}, $Z_{\odot} \Rightarrow 8.69=12+\log({\rm O/H})$.  
}
\end{table}

\section{Results}
\label{sec:discussion}
We summarize the main results in the following.

\begin{figure*}
\center
 \includegraphics[width=0.72\textwidth]{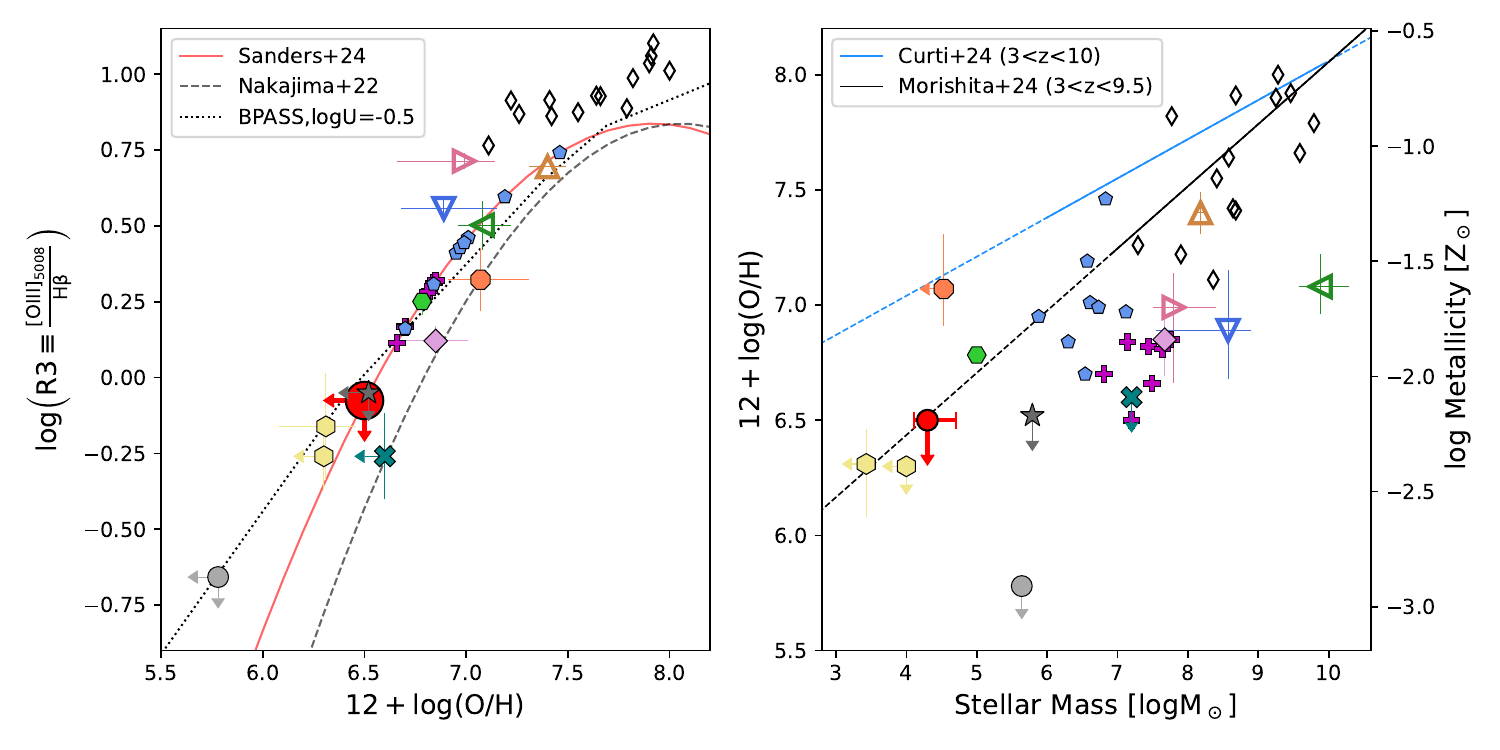}
 \includegraphics[width=0.24\textwidth]{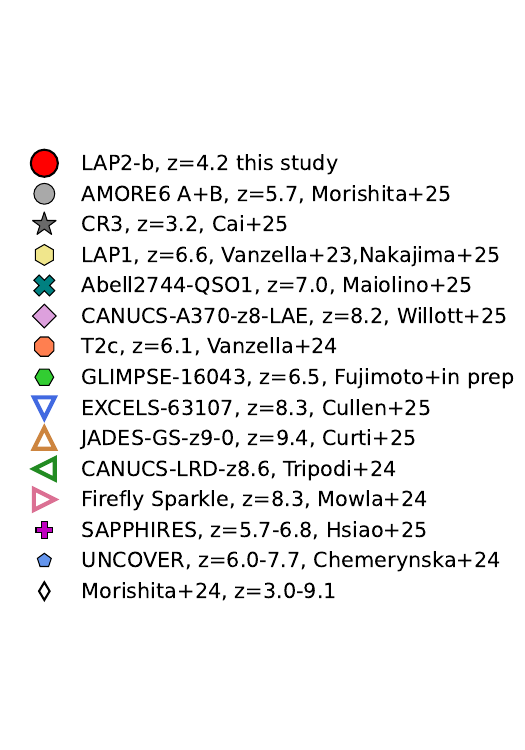}
 \caption{Oxygen abundance of LAP2-b, as function of its R3 index (left panel, 2$\sigma$ limit is reported) and its stellar mass (right panel).  Additional measurements of low-metallicity sources at $z>3$ from the literature are reported \citep{Vanzella2023_lap1,Morishita2024,Chemerynska2024,Mowla2024,Tripodi2025,Vanzella2024_t2c,Hsiao2025,Curti2025,Cullen2025,Willott2025,Maiolino2025,Nakajima2025,Cai2025,Morishita2025}. Samples plotted with empty markers have their metallicity estimated from the direct $T_e$ method, as opposed to the ones derived from strong line ratios (filled markers).
 The two most common R3-to-Z calibrations used in high-$z$ studies 
 as well as the conversion based on the BPASS models used in \citet{Morishita2025} (dotted black line), are shown in the left panel (see legend). Two recent mass-metallicity relations derived from high-z samples (blue line: \citealp{Curti2024}; black line: \citealp{Morishita2024}) are shown in the right panel; the mass ranges 
 used to derive the relations are shown as the solid portion of the line.
 }
 \label{fig:metal}
\end{figure*}

\noindent (1) Metallicity.  
The spectrum shows a very faint continuum with clear \lya\ and \ha\ emission lines, 
consistent with the VLT/MUSE \lya\ measured at 10$\sigma$ (Table~\ref{tab:properties}).
The most remarkable spectral feature is the non-detection of \oiiidoublam, down to a 2$\sigma$ upper limit of $0.50\times 10^{-19}$~\ergscm. This implies an R3 index $\lesssim 0.85$ (assuming the intrinsic \ha/\hb\ ratio of 2.86 from case~B recombination, \citealt{osterbrock2006}), which, assuming the R3-index calibration for high-z systems by \citet{Sanders2024}, corresponds to a metallicity of $Z < 0.006\,Z_{\odot}$ (or 12+log(O/H) $< 6.5$) at $2\sigma$. Even considering more conservative R3 calibrations \citep[e.g.,][]{nakajima2022}, the expected metallicity remains within $\sim1\%~Z_\odot$ (Figure~\ref{fig:metal})\footnote{The derived R3 limit would correspond to $\rm Z\gtrsim0.9~Z_\odot$ if considering the high-metallicity branch of the calibrations; in this case, we would expect detectable [\textrm{S}\textsc{ii}]\ensuremath{\lambda\lambda6717,6731}/\ha$\approx$0.4 and [\textrm{O}\textsc{ii}]\ensuremath{\lambda3727}/\hb$\approx$3, which we do not observe in the spectrum, rejecting this case.
}.

\noindent (2) Morphology.  
From the F090W-band image shown in Figure~\ref{fig:line}, LAP2-b appears point-like. We conservatively adopted the FWHM measured in the F090W band ($\simeq 30$~mas) as an estimate of its size. The tangential magnification of $\simeq 20 \pm 2$ corresponds to less than 10~pc along the tangential direction. A more detailed morphological analysis will be performed on the NIRCam images in combination with forthcoming NIRSpec/IFU observations of both mirrored images (LAP2-a and LAP2-b).

\noindent (3) Stellar mass and age.  
The F090W magnitude of LAP2-b is $m_{\mathrm{F090W}} = m_{\mathrm{UV}} = 29.69 \pm 0.09$ (probing $\lambda \simeq 1700$\AA~rest-frame), and with a magnification factor of $\mu = 50 \pm 5$ it translates into an intrinsic $m_{\mathrm{UV}} = 33.8$ or ultraviolet magnitude of $M_{\rm UV} \simeq  -12.2$ at $z=4.189$.  
The resulting equivalent width 
of the \ha\ line is 647~\AA\ when inferred from the \ha\ line flux and the F277W magnitude (or 600~\AA\ from the spectrum itself, though the continuum is poorly constrained; see Table~\ref{tab:properties}). The ionizing photon production efficiency $-$ the LyC production rate per monochromatic UV 
luminosity \citep[][]{schaerer16} $-$ is relatively high, Log$_{10}(\xi_{ion}) = 25.13$, and is a lower limit if the escaping ionizing radiation is not zero ($f_{\rm esc} > 0$, neglecting dust attenuation correction, as copious \lya\ emission suggests).
Given the compact size of the star complex, an instantaneous burst is a plausible star formation history. Using {\tt STARBURST99} models \citep[][]{leitherer14} with a bursty scenario and the lowest available metallicity ($= 1/20~Z_{\odot}$), the observed UV luminosity and the \ha\ equivalent width imply an age $<10$~Myr and a stellar mass $\simeq 2\times 10^4$~\msun. A similar range of stellar mass for the same interval of ages is inferred from the rest-frame optical wavelength (F277W, sampling the $\lambda \simeq 5000$ \AA, Tab.~\ref{tab:properties}), $(1.2-5.0)\times 10^4$~\msun.
This scenario also implies a significantly high specific star formation rate
of $\gtrsim 100$ Gyr$^{-1}$, corresponding to the inverse of the estimated age.
It is worth noting that the observed line ratio, \ha/\hb,\ is currently not constrained due to low S/R of
\hb\ detection ($ \simeq 2.8\sigma$), and the \lya/\ha\ (either from NIRSpec or in combination with MUSE) needs a dedicated NIRSpec IFU observation that fully captures the \lya\ and \ha\ emission regions (prog. 7677). 
Finally, LAP2-b appears 
rather 
isolated based on the current spectroscopic redshift catalogs. 
Using the DR4 spectroscopic 
redshift 
catalog released by the UNCOVER collaboration \citep[][]{Suess_2024}, we searched for galaxies within 10 cMpc of LAP2-b. 
We found only one galaxy with $z_\mathrm{spec}=4.181$, located 5.3 comoving Mpc from LAP2-b. 

\section{Final remarks}
The VLT/MUSE blind spectroscopy led to the discovery of an extremely faint star source at $z=4.19$ straddling the critical line \citep[][]{richard2021}, appearing as two mirrored images: LAP2-a and LAP2-b. The large magnification factors of the two images imply that
LAP2 is a tiny and faint star complex with an estimated size of $< 10$ pc, intrinsic ultraviolet luminosity $M_{\rm UV} =-12.2,$ and a stellar mass of a few tens of thousands of solar masses. The \JWST/NIRSpec spectrum shows a clear deficit of \oiiidoublam, $\sim 7$ times fainter than \ha\ 
corresponding to $2\sigma$ limit of $Z < 0.006$~$Z_{\odot}$. This places LAP2-b among the most metal poor objects known to date (Figure~\ref{fig:metal}) and re-opens the issue of prolonged pristine star formation down to relatively low redshifts: $z\sim 3-5$ \citep[e.g.,][]{Liu_Bromm_endPopIII2020}.
These sources provide an initial glimpse of galaxies under conditions similar to those of the early Universe, occupying a rarely explored regime of low stellar mass.
The system is scheduled to be observed with NIRSpec/IFU in prism mode (Prog.~7677) for a total of 17.4~hours. The resulting 2D maps will enable a direct comparison with the VLT/MUSE IFU \lya\ detection, an independent and deeper check of the oxygen deficit in both multiple images, and, by combining them with the data presented here, the possibility of placing more stringent limits on oxygen and other metal lines. These observations will push LAP2-b into a previously unexplored regime of very low-metallicity systems.

\begin{acknowledgements}
We thank the anonymous referees for the careful reading and constructive comments. 
This work is based on observations made with the NASA/ESA/CSA 
\textit{James Webb Space Telescope} (\JWST)
and \textit{Hubble Space Telescope} (\HST). 
These observations are associated with \JWST\ GO program n.2561 (PI I. Labb\'e). We acknowledge financial support through grants PRIN-MIUR 2017WSCC32 and 2020SKSTHZ. 
We thank S.Fujimoto for providing the data for GLIMPSE-16043. 
EV and MM acknowledge financial support through grants 
INAF GO Grant 2024 ``Mapping Star Cluster Feedback in a Galaxy 450 Myr after the Big Bang'' and the European Union – NextGenerationEU within PRIN 2022 project n.20229YBSAN - Globular clusters in cosmological simulations and lensed fields: from their birth to the present epoch. 
This research has used NASA’s Astrophysics Data System and SAOImageDS9, developed by Smithsonian Astrophysical Observatory.
Additionally, this work made use of the following open-source packages for Python, and we are thankful to the developers of these: Matplotlib \citep{matplotlib2007}, Numpy \citep[][]{NUMPY2011}. 
EV thanks Martini Urban Ristobar for the stimulating ideas and internet connection.
\end{acknowledgements}

\bibliographystyle{aa}
\bibliography{bib}

\end{document}